\documentclass[twocolumn,showpacs,preprintnumbers,prl,fleqn,superscriptaddress]{revtex4-2}
\usepackage[english]{babel}
\usepackage[utf8]{inputenc}
\usepackage{SIunits}
\usepackage{graphicx}
\usepackage{dcolumn}
\usepackage{amsmath}
\usepackage{natbib}
\usepackage{bm}
\usepackage{textcomp}
\usepackage{color}
\usepackage{float}
\usepackage{booktabs}
\usepackage{multirow}
\usepackage{tabularx}
\usepackage{adjustbox}
\usepackage{hyperref}
\usepackage{tikz}
\usepackage{makecell}

\begin{document}

\title{Probing Dielectric Screening in van der Waals Heterostructures via Pressure-Tuned Exciton Rydberg Series}

\author{Shalini Badola}
\affiliation{LNCMI, UPR 3228, CNRS, EMFL, Universit\'e Grenoble Alpes, 38000 Grenoble, France}
\author{Adlen Smiri}
\affiliation{Universit\'e de Toulouse, INSA-CNRS-UPS, LPCNO, 135 Avenue de Rangueil, 31077 Toulouse, France}
\author{Thomas Pelini}
\affiliation{LNCMI, UPR 3228, CNRS, EMFL, Universit\'e Grenoble Alpes, 38000 Grenoble, France}
\author{Aditi Moghe}
\affiliation{LNCMI, UPR 3228, CNRS, EMFL, Universit\'e Grenoble Alpes, 38000 Grenoble, France}
\author{Tristan Riccardi}
\affiliation{LNCMI, UPR 3228, CNRS, EMFL, Universit\'e Grenoble Alpes, 38000 Grenoble, France}
\author{Amit Pawbake}
\affiliation{LNCMI, UPR 3228, CNRS, EMFL, Universit\'e Grenoble Alpes, 38000 Grenoble, France}
\author{Kenji Watanabe}
\affiliation{Research Center for Electronic and Optical Materials, National Institute for Materials Science, 1-1 Namiki, Tsukuba 305-0044, Japan}
\author{Takashi Taniguchi}
\affiliation{Research Center for Materials Nanoarchitectonics, National Institute for Materials Science,  1-1 Namiki, Tsukuba 305-0044, Japan}
\author{Iann C. Gerber}
\affiliation{Universit\'e de Toulouse, INSA-CNRS-UPS, LPCNO, 135 Avenue de Rangueil, 31077 Toulouse, France}
\author{Cl\'ement Faugeras}
\email{clement.faugeras@lncmi.cnrs.fr}
\affiliation{LNCMI, UPR 3228, CNRS, EMFL, Universit\'e Grenoble Alpes, 38000 Grenoble, France}

\date{\today }

\begin{abstract}

Excitons in two-dimensional semiconductors are directly exposed to the environment and are sensitive to the dielectric properties of their surrounding. Here, we show that the Rydberg series of excited states of excitons in a monolayer WSe$_2$ encapsulated in hexagonal boron nitride (hBN) can be used to probe the pressure-induced modifications of the surrounding dielectric properties. We propose a model based on the pressure induced evolution of the interlayer distances in this van der Waals heterostructure and on the bulk dielectric properties of hBN. This approach allows a direct measurement of the dielectric constant of pressurized hBN and establishes a new methodology for dielectric sensing.

\end{abstract}

\maketitle

Excitons in two-dimensional (2D) semiconductors have attracted intense interest in recent years owing to their unusually large binding energies~\cite{Ugeda2014} and their ability to form a variety of many-body complexes, including trions~\cite{Mak2012}, biexcitons~\cite{You2015}, and dark excitons~\cite{molas2017}, endowed with coupled spin and valley degrees of freedom~\cite{Xu2014}. In monolayers (ML) of transition metal dichalcogenides (TMDs), reduced dielectric screening and quantum confinement lead to strongly bound excitonic states, whose series of excited states, the Rydberg series, has been observed even in samples deposited on SiO$_2$~\cite{chernikov2014,he2014}. The Rydberg series of excited states act as a fingerprint of the exciton dimensionality and dielectric environment~\cite{Yu2010}. However, systematic control of the Coulomb interaction and of the exciton Rydberg series has only become possible with the advent of high-quality van der Waals (vdW) heterostructures~\cite{Geim2013}.

Because of the nonlocal screening inherent to atomically thin materials, the Coulomb interaction in TMD monolayers is strongly influenced by the surrounding dielectric environment. As a result, modifying the encapsulation or substrate provides an efficient route to engineer the energy spacing of excited excitonic states~\cite{stier2016,Raja2017}. This behavior is commonly described by the Rytova–Keldysh potential~\cite{rytova1967,Keldysh1979}, which has regained considerable attention following the discovery of 2D semiconductors~\cite{Qiu2013,Wang2018,Dery2018,stier2018,Molas2019}. Remarkably, when a WSe$_2$ monolayer is encapsulated in hexagonal boron nitride (hBN) to improve its optical quality~\cite{Cadiz2017}, the resulting exciton Rydberg series closely approaches the hydrogenic $1/n^2$ scaling characteristic of three-dimensional excitons~\cite{stier2018,Molas2019}. In all these studies, however, the dielectric environment is fixed by the sample architecture and cannot be tuned in situ.

Hydrostatic pressure has recently emerged as a powerful and reversible tool to modify interlayer distances and interaction strengths in vdW materials and heterostructures. It has been successfully employed to enhance interlayer coupling in homo- and heterobilayers of TMDs~\cite{Hsu2022,Xia2020,Zhu2022}, to tune correlated states in twisted bilayer graphene~\cite{Yankowitz2019}, and to stabilize novel magnetic phases and quasiparticles in vdW magnets~\cite{Pawbake2023,Pawbake2022}. Despite these advances, the influence of pressure on dielectric screening in vdW heterostructures remains largely unexplored.

In this Letter, we exploit the sensitivity of the exciton Rydberg series in monolayer WSe$_2$ to probe pressure-induced modifications of the dielectric environment. Using low-temperature photoluminescence spectroscopy in a diamond anvil cell, we track the evolution of excited excitonic states in a high-quality hBN-encapsulated WSe$_2$ monolayer under hydrostatic pressure. We observe a systematic compression of the Rydberg series, revealing enhanced Coulomb screening. To interpret these results, we develop a microscopic dielectric model that accounts for the finite thickness of the monolayer and its separation from the surrounding hBN by an effective vdW gap. Our analysis demonstrates that the observed renormalization originates from the pressure-induced decrease of interlayer distances and hBN dielectric constant's increase.

\begin{figure}
\includegraphics[width=1\linewidth,angle=0,clip]{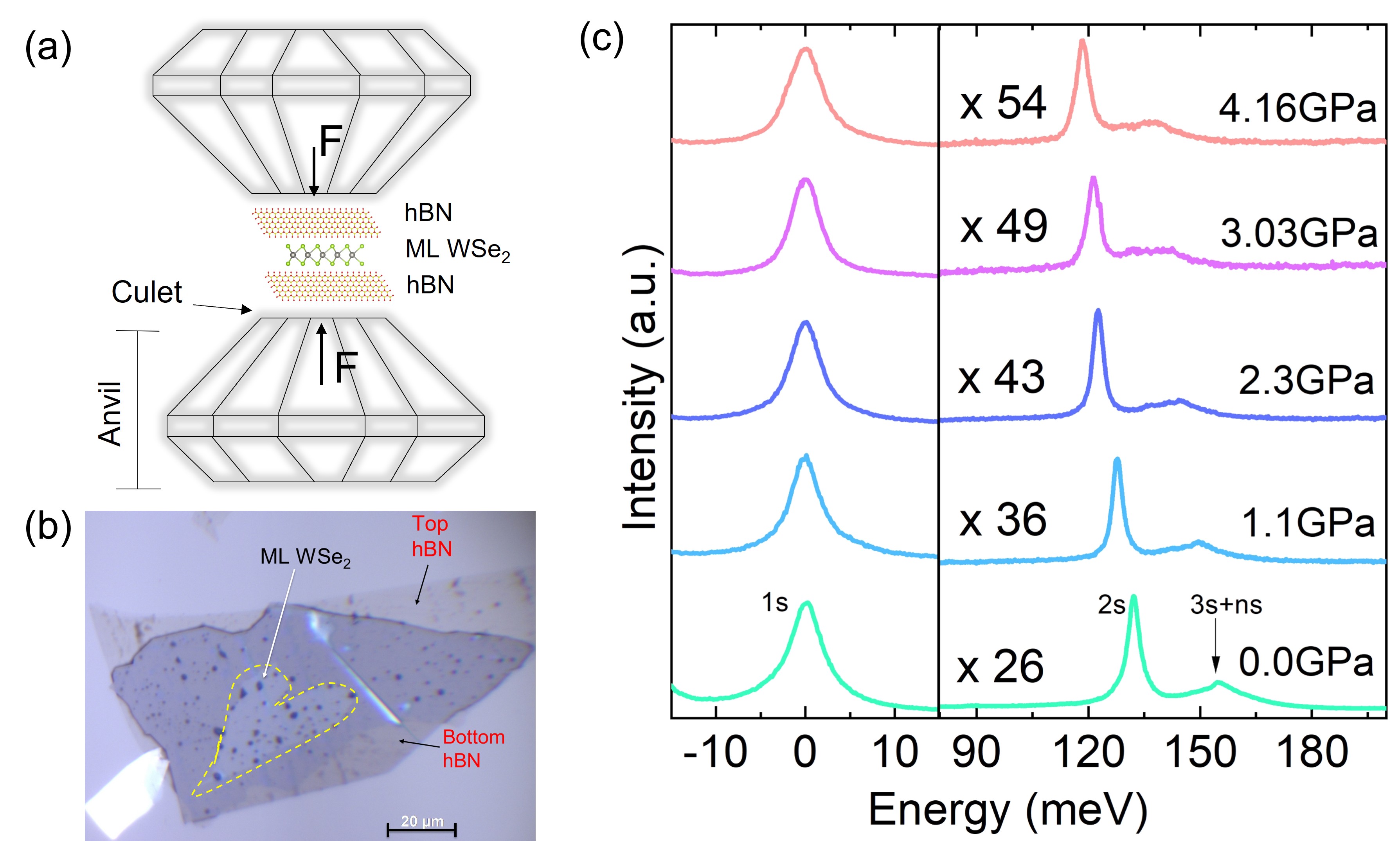}
\caption{a) Schematic of the experiment showing the diamond anvil cell with the vdW heterostructure exfoliated and transferred on the culet. The WSe$2$ ML is indicated with the yellow dashed line b) optical image of the vdW heterostructure on the culet. c) Pressure evolution of the photoluminescence for the encapsulated WSe$_2$ monolayer showing the modifications of the exciton Rydberg series. Spectra are shifted in energy to the $1s$ state energy for each pressure, being normalized with respect to the $1s$ state intensity and the excited states being scaled with the factor indicated for each pressure. $ns$ denotes exciton excited states with principal quantum number higher than 3.
\label{Fig1}}
\end{figure}

Thin layers of hBN and a monolayer of WSe$_2$ have been exfoliated from bulk material using scotch-tape technique and the hBN/WSe$_2$/hBN heterostructure has been assembled on the culet of a diamond anvil cell (DAC) using PDMS-based dry transfer technique. Figure~\ref{Fig1}a shows a schematic of the DAC with the vdW heterostructure transferred by PDMS dry-transfer technique~\cite{gomez2014,Goossens2012,Kim2019} on one of the culets of the DAC. An optical image of the vdW heterostructure comprising a bottom $38$~nm hBN layer, a monolayer WSe$_2$ and then the top $10$~nm hBN layer is presented in Figure~\ref{Fig1}b. We use ethanol-methanol (4:1) as pressure transmitting medium and the pressure value is determined with the luminescence from a ruby ball inserted in the pressure chamber. Additional evidence for the effective application of pressure is provided by the Raman scattering response of WSe$_2$ under pressure presented in the Supplemental Informations. The DAC is then placed on piezo motors and inserted into a set-up allowing for magneto-optical investigations at low temperature and under high hydrostatic pressure~\cite{Breslavetz2021}.

When reducing temperature down to liquid helium temperature, the exciton Rydberg series appears in the photoluminescence spectrum, see Figure~\ref{Fig1}c. In ML WSe$_2$, one typically observes the $1s$, the $2s$ states well spectrally separated, and a broader PL feature at higher energy which was shown to comprise the $3s$ state together with higher index excited states. When applying a magnetic field perpendicular to the WSe$_2$ layer, the exciton states evolve in energy as has been already reported~\cite{stier2018,Molas2019}, proving that these PL features do arise from exciton, see supplemental material. Applying hydrostatic pressure to this vdW heterostructure, we observe a gradual change of the exciton excited states Rydberg series: the energy separation between the different excited states varies, see Fig.~\ref{Fig1}c. Because of spatial inhomogeneities in the sample, the absolute $1s$ state energy varies slightly depending on the location on the sample. For this reason, we take the $1s$ state energy as the energy reference and we evaluate the energy separation between a given excited state and this value. Thus hydrostatic pressure does influence the Rydberg series, and hypothetically by two main origins. Either pressure induces a modification of the electrostatic screening, and/or modify sufficiently the electronic structure of the TMD ML to affect the exciton reduced mass.

\begin{figure}
\includegraphics[angle=0,width=0.5\textwidth]{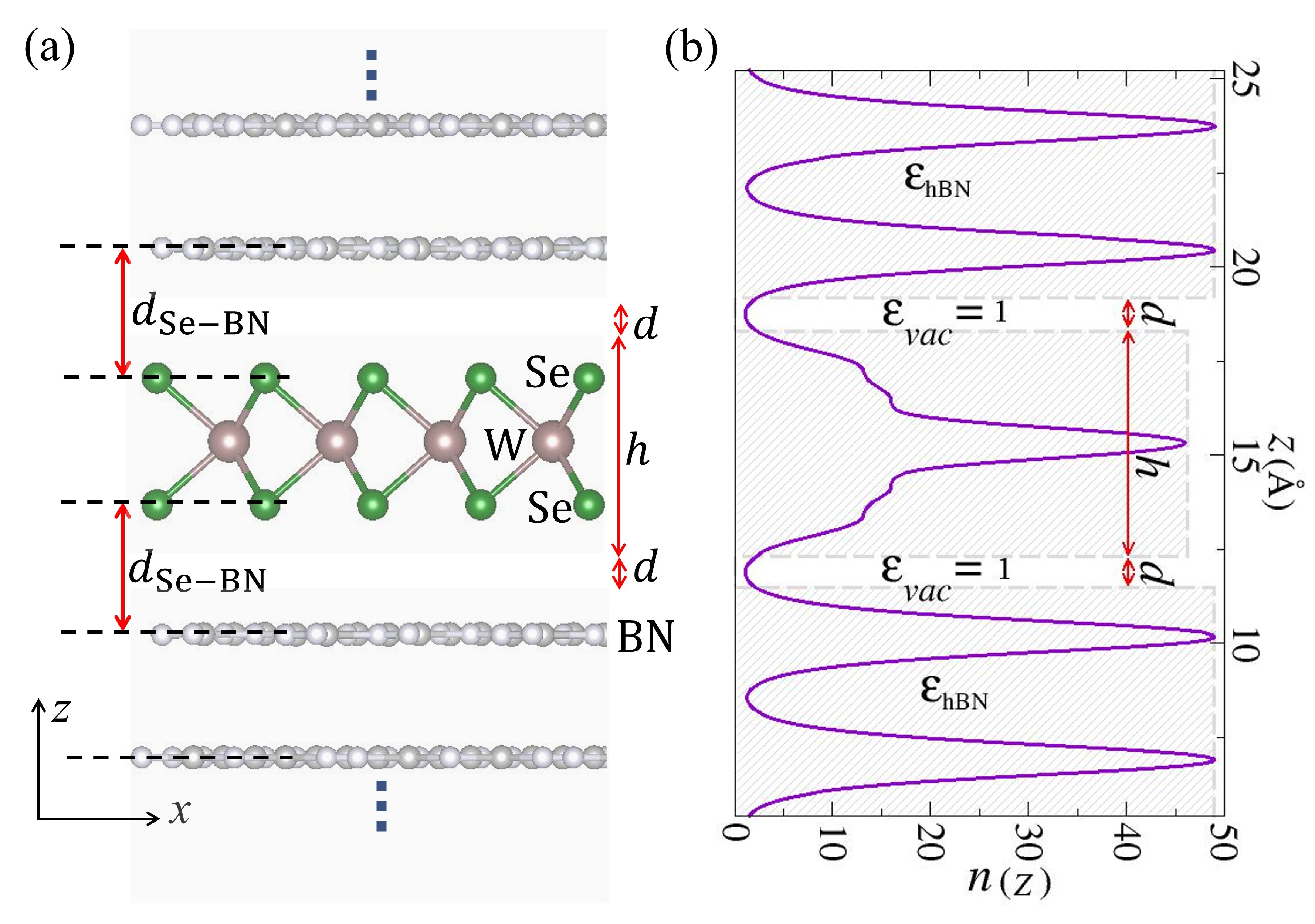}
\caption{a) Atomic structure of the WSe$_2$ ML encapsulated in hBN. b) Planar-averaged electron density $n(z)$ as a function of the out-of-plane coordinate $z$. The profile reveals the electron distribution across the slab, with pronounced peaks at the atomic layers and a minimum in the interlayer region. The electron density $n(z)$ near the minimum is used to define the boundaries of the interlayer region. Here, $d$ and $h$ are the interlayer distance and the WSe$_2$ thickness, respectively.
\label{Fig2}}
\end{figure}

To discriminate between the two effects, we have developed the following model, based on density functional theory (DFT) calculations, see Supplemental Informations for computational details, in which we consider a WSe$_2$ ML of thickness $h$ sandwiched between two bulk hBN layers as sketched in Fig.~\ref{Fig2}a. The out-of-plane charge density distribution $n(z)$ is shown in Fig.~\ref{Fig2}b. The function $n(z)$ is obtained by averaging the \textit{ab initio} three-dimensional charge density over the $xy$ plane, as described in the supplemental material. Within the WSe$_2$ layer, $n(z)$ is approximated by a step function with amplitude $h$, as illustrated by the dashed profile in Fig.~\ref{Fig2}b. A vacuum-like gap of width $d$ separates the ML from the surrounding hBN. In this gap, the charge density is negligible and the dielectric constant is taken to be that of free space. Within this frame, the electron–hole interaction is described by the quasi-2D potential of Ref.~\cite{Latini2015}:
\begin{equation}
V_{\rm SC}(\rho)=-\frac{2e^{2}}{h}\int dq\,\frac{J_{0}(\rho q)}{q\,\varepsilon_{\rm SC}(q)}\left[1-\frac{2}{hq}e^{-hq/2}\sinh\biggl{(}\frac{qh}{2}\biggr{)}\right]
\end{equation}

Here, $J_{0}(x)$ denotes the zeroth-order Bessel function, and $h$ being the thickness of the WSe$_2$ ML, which characterizes the extent of the out-of-plane charge density distribution. The dielectric function $\varepsilon_{SC}(q)$ is given by~\cite{Florian2018}:

\begin{equation}
\varepsilon_{\rm SC}(q) = \varepsilon_{\rm WSe_{2}}(q)\,\frac{1-\epsilon_{1}e^{-q(h+d)} - \epsilon_{2}e^{-qh} + \epsilon_{1}\epsilon_{2}e^{-qd}}{1 + \epsilon_{1}e^{-q(h+d)} + \epsilon_{2}e^{-qh} + \epsilon_{1}\epsilon_{2}e^{-qd}}
\end{equation}
where $ \epsilon_{1} = \displaystyle \frac{\varepsilon_{\rm vac} - \varepsilon_{\rm hBN}}{\varepsilon_{\rm vac} + \varepsilon_{\rm hBN}}$ and $\varepsilon_{2}(q) = \displaystyle \frac{\varepsilon_{\rm WSe_{2}}(q) - \varepsilon_{\rm hBN}}{\varepsilon_{\rm WSe_{2}}(q) + \varepsilon_{\rm hBN}}$ with $\varepsilon_{\rm vac} = 1$. Here, $\varepsilon_{\rm hBN}$ and $\varepsilon_{\rm vac}$ denote the dielectric constants of the bulk hBN and the interlayer gap, respectively. The in-plane momentum-dependent dielectric function $\varepsilon_{\rm WSe_2}(q)$ is the dielectric response of the WSe$_2$ ML, neglecting nonlocal effects along the $z$-direction~\cite{Florian2018,rosner2015}.

\begin{figure}
\includegraphics[width=1\linewidth,angle=0,clip]{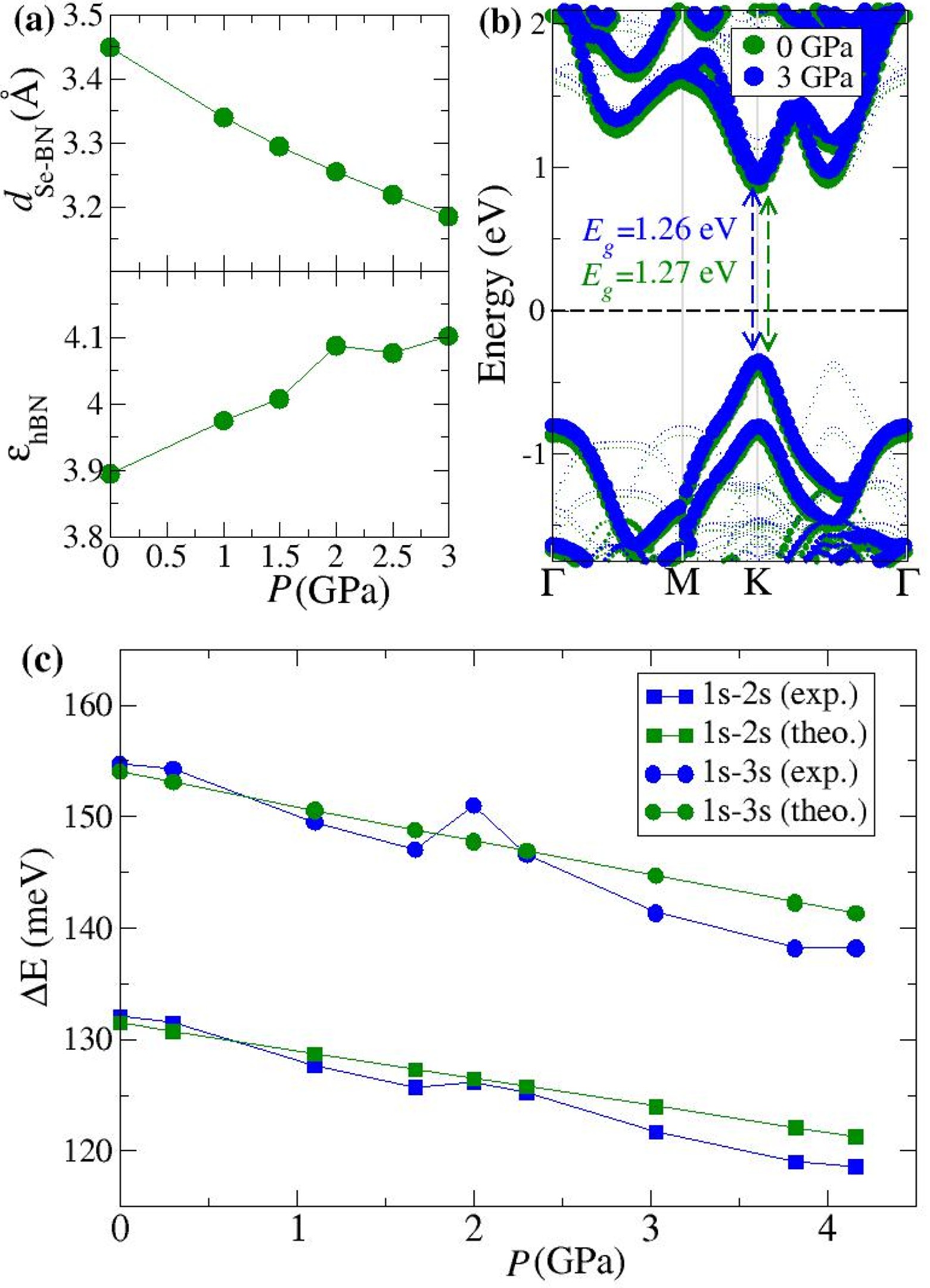}
\caption{a) Dielectric constant ($\epsilon_{\rm{hBN}}$) of the bulk hBN material and interlayer distance d$_{\rm Se-BN}$ as a function of hydrostatic pressure $P$.  b) Electronic band structure at $P=$ 0 (green) and at $P=$ 3 GPa (blue). c) Comparison between experimental data and theoretical predictions for the pressure dependence of the energy separations $\Delta$E$_{1s-2s}$ and $\Delta$E$_{1s-3s}$ of excitonic states in hBN-encapsulated WSe$_2$ ML as a function of hydrostatic pressure.
\label{Fig3}}
\end{figure}

Hydrostatic pressure is introduced by considering the pressure dependence of the dielectric function through two quantities: the interlayer distance $d(P)$ and the bulk dielectric constant of hBN, $\varepsilon_{\rm hBN}(P )$. Both of them are obtained from DFT calculations and presented in Figure~\ref{Fig3}a. This distance decreases from $3.45$~\AA~at ambient pressure down to $3.18$~\AA~for a pressure of $P=3$~GPa. See Supplemental Informations for more computational details~\cite{Kresse1993,Kresse1996,Perdew1996,Blochl1994,Kresse1999,Klime2009,Wang2021}.

In Ref.~[\citenum{Florian2018}], the interlayer vacuum spacing is identified as the structural interlayer distance, $d = d_{\rm Se-BN}$. However, as shown in Fig.~\ref{Fig2}b, the extended electron density in the interlayer region implies a smaller value of $d$. In fact, the electronic density is negligible only in a region of about $1$~\AA~height, supporting the assumption of an effective vacuum spacing. The value of $d$ at ambient-pressure can be estimated by fitting the experimental exciton energies at $P=0$. A value of $d=0.8$~\AA~in the model reproduces well both the $1s$ exciton energy and the energy separations between the $1s-2s$ and $1s-3s$ excitonic states when compared to experiments. Therefore, in the following we keep $d=0.8$~\AA~as reference value $d(P = 0)$.

The pressure dependence of the interlayer vacuum spacing $d(P)$ cannot be accessed directly from DFT. We therefore assume that the relative variation of $d(P)$ with pressure is the same as that of the Se–BN separation, i.e., $d(P)=d(P=0)[1-\kappa P]$, using the same $\kappa$ value as in $d_{\rm Se-BN}(P)= d_{\rm Se-BN}(P=0)[1-\kappa P]$. The pressure evolution of the interlayer distance $d_{\rm Se-BN}$ is, on the contrary, accessible by DFT calculations, and allows to estimate $\kappa$ from data given in Figure.~\ref{Fig3}a.

After modeling the impact of dielectric environment changes, we also computed the electronic band structure of WSe$_2$ at the standard DFT level, both in the unpressurized case and at $P = 3$~GPa. Our calculations show that the band structure of monolayer WSe$_2$ is only marginally affected by a pressure of $P = 3$~GPa, with a negligible shift in the fundamental band gap of approximately $10$~meV (Fig.~\ref{Fig3}b), and no discernible modification of the effective masses of the valence and conduction bands at the K point. This finding is corroborated by two magneto-optical experiments. Magneto-photoluminescence measurements reported in the supplemental material at ambient pressure, $P = 2$~GPa, and at $P = 4$~GPa yield a pressure-independent excitonic g-factor of $-4.3 \pm 0.1$, confirming that the electronic band structure remains largely insensitive to hydrostatic pressure~\cite{Rybkovskiy2017}. Furthermore, we measured the energies of the $1s$ excitonic states arising from different bands via magnetic brightening~\cite{molas2017}, in order to track the pressure dependence of the two lowest conduction bands. The results presented in the supplemental material reveal that the dark-bright energy splitting varies only slightly with pressure, remaining below $3$~meV at $P = 3$~GPa. This further confirms that the electronic band structure is only weakly perturbed under pressure, such that the exciton reduced mass remains nearly constant and the exciton Rydberg energy should only be marginally affected.


The pressure-induced changes in the dielectric properties of the WSe$_2$ ML environment is therefore the primary cause of the shifts in the excitonic Rydberg series. We compare in Figure~\ref{Fig3}c the experimental $E_{1s-2s}$ and $E_{1s-3s}$, characteristic of the Rydberg ladder, with those calculated considering pressure induced changes of the dielectric properties of the whole vdW heterostructure's dielectric properties. The agreement between these different sets of data shows that excitons in ML TMDs are sensitive to pressure-induced modifications of the dielectric environment and that they can be used to sense such changes. As pressure increases, we observe a reduction in the luminescence intensity of excited states relative to the 1s state, see Fig.~\ref{Fig1}c. This trend is consistent with a pressure-enhanced carrier relaxation mechanism, whereby the shrinking energy gap between excited and ground states, induced by stronger screening, facilitates faster depopulation of the excited states.

In conclusion, we have demonstrated that the exciton Rydberg series in a hBN-encapsulated ML WSe$_2$ provides a sensitive and quantitative probe of pressure-induced modifications of dielectric screening in vdW heterostructures. By performing low-temperature photoluminescence spectroscopy under hydrostatic pressure, we directly track the evolution of excited excitonic states and reveal a systematic compression of the Rydberg ladder, indicative of enhanced Coulomb screening.

Using a microscopic dielectric model that accounts for the finite thickness of the ML and its separation from the surrounding hBN by a vdW gap, we establish that the observed renormalization originates predominantly from the pressure-induced increase of the hBN dielectric constant. This approach enables an optical determination of dielectric properties under extreme conditions with nanoscale spatial sensitivity.

Beyond the specific case of hBN/WSe$_2$/hBN heterostructures, our results highlight the broader potential of excitons in two-dimensional semiconductors as quantum sensors of their electrostatic environment~\cite{Popert2022,He2024}. This methodology can be readily extended to other layered materials~\cite{Castellanos2011,hay2026}, heterostructure geometries, and external stimuli, providing a general framework for Coulomb engineering and dielectric metrology in low-dimensional systems. Our approach establishes excitons in 2D semiconductors as versatile quantum sensors for dielectric properties under extreme conditions, with potential applications in Coulomb engineering and the design of optoelectronic devices.

\begin{acknowledgements}
This work was partially supported by LNCMI, a member of the European Magnetic Field Laboratory (EMFL). This work is also partially supported by ANR-23-QUAC-0004, and by CEFIPRA CSRP Project No.~7104-2. This work is partially supported by France 2030 government investment plan managed by the French National Research Agency under grant reference PEPR SPIN - SPINMAT ANR- 22-EXSP-0007. I. C. G. and A. S. gratefully acknowledge the computational resources provided by the CALMIP initiative (project P0812) and CINES, IDRIS, and TGCC, which were granted by GENCI through allocation 2025-A0180906649. K.W. and T.T. acknowledge support from the JSPS KAKENHI (Grant Numbers 21H05233 and 23H02052), the CREST (JPMJCR24A5), JST and World Premier International Research Center Initiative (WPI), MEXT, Japan.
\end{acknowledgements}


\begin{thebibliography}{44}%
\makeatletter
\providecommand \@ifxundefined [1]{%
 \@ifx{#1\undefined}
}%
\providecommand \@ifnum [1]{%
 \ifnum #1\expandafter \@firstoftwo
 \else \expandafter \@secondoftwo
 \fi
}%
\providecommand \@ifx [1]{%
 \ifx #1\expandafter \@firstoftwo
 \else \expandafter \@secondoftwo
 \fi
}%
\providecommand \natexlab [1]{#1}%
\providecommand \enquote  [1]{``#1''}%
\providecommand \bibnamefont  [1]{#1}%
\providecommand \bibfnamefont [1]{#1}%
\providecommand \citenamefont [1]{#1}%
\providecommand \href@noop [0]{\@secondoftwo}%
\providecommand \href [0]{\begingroup \@sanitize@url \@href}%
\providecommand \@href[1]{\@@startlink{#1}\@@href}%
\providecommand \@@href[1]{\endgroup#1\@@endlink}%
\providecommand \@sanitize@url [0]{\catcode `\\12\catcode `\$12\catcode
  `\&12\catcode `\#12\catcode `\^12\catcode `\_12\catcode `\%12\relax}%
\providecommand \@@startlink[1]{}%
\providecommand \@@endlink[0]{}%
\providecommand \url  [0]{\begingroup\@sanitize@url \@url }%
\providecommand \@url [1]{\endgroup\@href {#1}{\urlprefix }}%
\providecommand \urlprefix  [0]{URL }%
\providecommand \Eprint [0]{\href }%
\providecommand \doibase [0]{https://doi.org/}%
\providecommand \selectlanguage [0]{\@gobble}%
\providecommand \bibinfo  [0]{\@secondoftwo}%
\providecommand \bibfield  [0]{\@secondoftwo}%
\providecommand \translation [1]{[#1]}%
\providecommand \BibitemOpen [0]{}%
\providecommand \bibitemStop [0]{}%
\providecommand \bibitemNoStop [0]{.\EOS\space}%
\providecommand \EOS [0]{\spacefactor3000\relax}%
\providecommand \BibitemShut  [1]{\csname bibitem#1\endcsname}%
\let\auto@bib@innerbib\@empty
\bibitem [{\citenamefont {Ugeda}\ \emph {et~al.}(2014)\citenamefont {Ugeda},
  \citenamefont {Bradley}, \citenamefont {Shi}, \citenamefont {da~Jornada},
  \citenamefont {Zhang}, \citenamefont {Qiu}, \citenamefont {Ruan},
  \citenamefont {Mo}, \citenamefont {Hussain}, \citenamefont {Shen},
  \citenamefont {Wang}, \citenamefont {Louie},\ and\ \citenamefont
  {Crommie}}]{Ugeda2014}%
  \BibitemOpen
  \bibfield  {author} {\bibinfo {author} {\bibfnamefont {M.~M.}\ \bibnamefont
  {Ugeda}}, \bibinfo {author} {\bibfnamefont {A.~J.}\ \bibnamefont {Bradley}},
  \bibinfo {author} {\bibfnamefont {S.-F.}\ \bibnamefont {Shi}}, \bibinfo
  {author} {\bibfnamefont {F.~H.}\ \bibnamefont {da~Jornada}}, \bibinfo
  {author} {\bibfnamefont {Y.}~\bibnamefont {Zhang}}, \bibinfo {author}
  {\bibfnamefont {D.~Y.}\ \bibnamefont {Qiu}}, \bibinfo {author} {\bibfnamefont
  {W.}~\bibnamefont {Ruan}}, \bibinfo {author} {\bibfnamefont {S.-K.}\
  \bibnamefont {Mo}}, \bibinfo {author} {\bibfnamefont {Z.}~\bibnamefont
  {Hussain}}, \bibinfo {author} {\bibfnamefont {Z.-X.}\ \bibnamefont {Shen}},
  \bibinfo {author} {\bibfnamefont {F.}~\bibnamefont {Wang}}, \bibinfo {author}
  {\bibfnamefont {S.~G.}\ \bibnamefont {Louie}},\ and\ \bibinfo {author}
  {\bibfnamefont {M.~F.}\ \bibnamefont {Crommie}},\ }\bibfield  {title}
  {\bibinfo {title} {Giant bandgap renormalization and excitonic effects in a
  monolayer transition metal dichalcogenide semiconductor},\ }\href
  {http://dx.doi.org/10.1038/nmat4061} {\bibfield  {journal} {\bibinfo
  {journal} {Nat. Mater.}\ }\textbf {\bibinfo {volume} {13}},\ \bibinfo {pages}
  {1091} (\bibinfo {year} {2014})}\BibitemShut {NoStop}%
\bibitem [{\citenamefont {Mak}\ \emph {et~al.}(2012)\citenamefont {Mak},
  \citenamefont {He}, \citenamefont {Shan},\ and\ \citenamefont
  {Heinz}}]{Mak2012}%
  \BibitemOpen
  \bibfield  {author} {\bibinfo {author} {\bibfnamefont {K.~F.}\ \bibnamefont
  {Mak}}, \bibinfo {author} {\bibfnamefont {K.}~\bibnamefont {He}}, \bibinfo
  {author} {\bibfnamefont {J.}~\bibnamefont {Shan}},\ and\ \bibinfo {author}
  {\bibfnamefont {T.~F.}\ \bibnamefont {Heinz}},\ }\bibfield  {title} {\bibinfo
  {title} {Control of valley polarization in monolayer mos2 by optical
  helicity},\ }\href {https://doi.org/10.1038/nnano.2012.96} {\bibfield
  {journal} {\bibinfo  {journal} {Nat. Nanotechnol.}\ }\textbf {\bibinfo
  {volume} {7}},\ \bibinfo {pages} {494} (\bibinfo {year} {2012})}\BibitemShut
  {NoStop}%
\bibitem [{\citenamefont {You}\ \emph {et~al.}(2015)\citenamefont {You},
  \citenamefont {Zhang}, \citenamefont {Berkelbach}, \citenamefont {Hybertsen},
  \citenamefont {Reichman},\ and\ \citenamefont {Heinz}}]{You2015}%
  \BibitemOpen
  \bibfield  {author} {\bibinfo {author} {\bibfnamefont {Y.}~\bibnamefont
  {You}}, \bibinfo {author} {\bibfnamefont {X.-X.}\ \bibnamefont {Zhang}},
  \bibinfo {author} {\bibfnamefont {T.~C.}\ \bibnamefont {Berkelbach}},
  \bibinfo {author} {\bibfnamefont {M.~S.}\ \bibnamefont {Hybertsen}}, \bibinfo
  {author} {\bibfnamefont {D.~R.}\ \bibnamefont {Reichman}},\ and\ \bibinfo
  {author} {\bibfnamefont {T.~F.}\ \bibnamefont {Heinz}},\ }\bibfield  {title}
  {\bibinfo {title} {Observation of biexcitons in monolayer wse2},\ }\href
  {https://doi.org/10.1038/nphys3324} {\bibfield  {journal} {\bibinfo
  {journal} {Nat. Phys.}\ }\textbf {\bibinfo {volume} {11}},\ \bibinfo {pages}
  {477} (\bibinfo {year} {2015})}\BibitemShut {NoStop}%
\bibitem [{\citenamefont {Molas}\ \emph {et~al.}(2017)\citenamefont {Molas},
  \citenamefont {Faugeras}, \citenamefont {Slobodeniuk}, \citenamefont
  {Nogajewski}, \citenamefont {Bartos}, \citenamefont {Basko},\ and\
  \citenamefont {Potemski}}]{molas2017}%
  \BibitemOpen
  \bibfield  {author} {\bibinfo {author} {\bibfnamefont {M.~R.}\ \bibnamefont
  {Molas}}, \bibinfo {author} {\bibfnamefont {C.}~\bibnamefont {Faugeras}},
  \bibinfo {author} {\bibfnamefont {A.~O.}\ \bibnamefont {Slobodeniuk}},
  \bibinfo {author} {\bibfnamefont {K.}~\bibnamefont {Nogajewski}}, \bibinfo
  {author} {\bibfnamefont {M.}~\bibnamefont {Bartos}}, \bibinfo {author}
  {\bibfnamefont {D.~M.}\ \bibnamefont {Basko}},\ and\ \bibinfo {author}
  {\bibfnamefont {M.}~\bibnamefont {Potemski}},\ }\bibfield  {title} {\bibinfo
  {title} {Brightening of dark excitons in monolayers of semiconducting
  transition metal dichalcogenides},\ }\href
  {https://doi.org/10.1088/2053-1583/aa5521} {\bibfield  {journal} {\bibinfo
  {journal} {2D Materials}\ }\textbf {\bibinfo {volume} {4}},\ \bibinfo {pages}
  {021003} (\bibinfo {year} {2017})}\BibitemShut {NoStop}%
\bibitem [{\citenamefont {Xu}\ \emph {et~al.}(2014)\citenamefont {Xu},
  \citenamefont {Yao}, \citenamefont {Xiao},\ and\ \citenamefont
  {Heinz}}]{Xu2014}%
  \BibitemOpen
  \bibfield  {author} {\bibinfo {author} {\bibfnamefont {X.}~\bibnamefont
  {Xu}}, \bibinfo {author} {\bibfnamefont {W.}~\bibnamefont {Yao}}, \bibinfo
  {author} {\bibfnamefont {D.}~\bibnamefont {Xiao}},\ and\ \bibinfo {author}
  {\bibfnamefont {T.~F.}\ \bibnamefont {Heinz}},\ }\bibfield  {title} {\bibinfo
  {title} {Spin and pseudospins in layered transition metal dichalcogenides},\
  }\href {https://doi.org/10.1038/nphys2942} {\bibfield  {journal} {\bibinfo
  {journal} {Nature Physics}\ }\textbf {\bibinfo {volume} {10}},\ \bibinfo
  {pages} {343–350} (\bibinfo {year} {2014})}\BibitemShut {NoStop}%
\bibitem [{\citenamefont {Chernikov}\ \emph {et~al.}(2014)\citenamefont
  {Chernikov}, \citenamefont {Berkelbach}, \citenamefont {Hill}, \citenamefont
  {Rigosi}, \citenamefont {Li}, \citenamefont {Aslan}, \citenamefont
  {Reichman}, \citenamefont {Hybertsen},\ and\ \citenamefont
  {Heinz}}]{chernikov2014}%
  \BibitemOpen
  \bibfield  {author} {\bibinfo {author} {\bibfnamefont {A.}~\bibnamefont
  {Chernikov}}, \bibinfo {author} {\bibfnamefont {T.~C.}\ \bibnamefont
  {Berkelbach}}, \bibinfo {author} {\bibfnamefont {H.~M.}\ \bibnamefont
  {Hill}}, \bibinfo {author} {\bibfnamefont {A.}~\bibnamefont {Rigosi}},
  \bibinfo {author} {\bibfnamefont {Y.}~\bibnamefont {Li}}, \bibinfo {author}
  {\bibfnamefont {O.~B.}\ \bibnamefont {Aslan}}, \bibinfo {author}
  {\bibfnamefont {D.~R.}\ \bibnamefont {Reichman}}, \bibinfo {author}
  {\bibfnamefont {M.~S.}\ \bibnamefont {Hybertsen}},\ and\ \bibinfo {author}
  {\bibfnamefont {T.~F.}\ \bibnamefont {Heinz}},\ }\bibfield  {title} {\bibinfo
  {title} {{Exciton Binding Energy and Nonhydrogenic Rydberg Series in
  Monolayer ${\mathrm{WS}}_{2}$}},\ }\href
  {https://doi.org/10.1103/PhysRevLett.113.076802} {\bibfield  {journal}
  {\bibinfo  {journal} {Phys. Rev. Lett.}\ }\textbf {\bibinfo {volume} {113}},\
  \bibinfo {pages} {076802} (\bibinfo {year} {2014})}\BibitemShut {NoStop}%
\bibitem [{\citenamefont {He}\ \emph {et~al.}(2014)\citenamefont {He},
  \citenamefont {Kumar}, \citenamefont {Zhao}, \citenamefont {Wang},
  \citenamefont {Mak}, \citenamefont {Zhao},\ and\ \citenamefont
  {Shan}}]{he2014}%
  \BibitemOpen
  \bibfield  {author} {\bibinfo {author} {\bibfnamefont {K.}~\bibnamefont
  {He}}, \bibinfo {author} {\bibfnamefont {N.}~\bibnamefont {Kumar}}, \bibinfo
  {author} {\bibfnamefont {L.}~\bibnamefont {Zhao}}, \bibinfo {author}
  {\bibfnamefont {Z.}~\bibnamefont {Wang}}, \bibinfo {author} {\bibfnamefont
  {K.~F.}\ \bibnamefont {Mak}}, \bibinfo {author} {\bibfnamefont
  {H.}~\bibnamefont {Zhao}},\ and\ \bibinfo {author} {\bibfnamefont
  {J.}~\bibnamefont {Shan}},\ }\bibfield  {title} {\bibinfo {title} {{Tightly
  Bound Excitons in Monolayer ${\mathrm{WSe}}_{2}$}},\ }\href
  {https://doi.org/10.1103/PhysRevLett.113.026803} {\bibfield  {journal}
  {\bibinfo  {journal} {Phys. Rev. Lett.}\ }\textbf {\bibinfo {volume} {113}},\
  \bibinfo {pages} {026803} (\bibinfo {year} {2014})}\BibitemShut {NoStop}%
\bibitem [{\citenamefont {Yu}\ and\ \citenamefont {Cardona}(2010)}]{Yu2010}%
  \BibitemOpen
  \bibfield  {author} {\bibinfo {author} {\bibfnamefont {P.~Y.}\ \bibnamefont
  {Yu}}\ and\ \bibinfo {author} {\bibfnamefont {M.}~\bibnamefont {Cardona}},\
  }\href {https://doi.org/10.1007/978-3-642-00710-1} {\emph {\bibinfo {title}
  {Fundamentals of Semiconductors: Physics and Materials Properties}}}\
  (\bibinfo  {publisher} {Springer Berlin Heidelberg},\ \bibinfo {year}
  {2010})\BibitemShut {NoStop}%
\bibitem [{\citenamefont {Geim}\ and\ \citenamefont
  {Grigorieva}(2013)}]{Geim2013}%
  \BibitemOpen
  \bibfield  {author} {\bibinfo {author} {\bibfnamefont {A.~K.}\ \bibnamefont
  {Geim}}\ and\ \bibinfo {author} {\bibfnamefont {I.}~\bibnamefont
  {Grigorieva}},\ }\bibfield  {title} {\bibinfo {title} {Van der waals
  heterostructures},\ }\href {https://www.nature.com/articles/nature12385}
  {\bibfield  {journal} {\bibinfo  {journal} {Nature}\ }\textbf {\bibinfo
  {volume} {499}},\ \bibinfo {pages} {419} (\bibinfo {year}
  {2013})}\BibitemShut {NoStop}%
\bibitem [{\citenamefont {Stier}\ \emph {et~al.}(2016)\citenamefont {Stier},
  \citenamefont {Wilson}, \citenamefont {Clark}, \citenamefont {Xu},\ and\
  \citenamefont {Crooker}}]{stier2016}%
  \BibitemOpen
  \bibfield  {author} {\bibinfo {author} {\bibfnamefont {A.~V.}\ \bibnamefont
  {Stier}}, \bibinfo {author} {\bibfnamefont {N.~P.}\ \bibnamefont {Wilson}},
  \bibinfo {author} {\bibfnamefont {G.}~\bibnamefont {Clark}}, \bibinfo
  {author} {\bibfnamefont {X.}~\bibnamefont {Xu}},\ and\ \bibinfo {author}
  {\bibfnamefont {S.~A.}\ \bibnamefont {Crooker}},\ }\bibfield  {title}
  {\bibinfo {title} {Probing the influence of dielectric environment on
  excitons in monolayer ${\mathrm{wse}}_{2}$: Insight from high magnetic
  fields},\ }\href {https://doi.org/10.1021/acs.nanolett.6b03276} {\bibfield
  {journal} {\bibinfo  {journal} {Nano Letters}\ }\textbf {\bibinfo {volume}
  {16}},\ \bibinfo {pages} {7054} (\bibinfo {year} {2016})}\BibitemShut
  {NoStop}%
\bibitem [{\citenamefont {Raja}\ \emph {et~al.}(2017)\citenamefont {Raja},
  \citenamefont {Chaves}, \citenamefont {Yu}, \citenamefont {Arefe},
  \citenamefont {Hill}, \citenamefont {Rigosi}, \citenamefont {Berkelbach},
  \citenamefont {Nagler}, \citenamefont {Sch{\"u}ller}, \citenamefont {Korn},
  \citenamefont {Nuckolls}, \citenamefont {Hone}, \citenamefont {Brus},
  \citenamefont {Heinz}, \citenamefont {Reichman},\ and\ \citenamefont
  {Chernikov}}]{Raja2017}%
  \BibitemOpen
  \bibfield  {author} {\bibinfo {author} {\bibfnamefont {A.}~\bibnamefont
  {Raja}}, \bibinfo {author} {\bibfnamefont {A.}~\bibnamefont {Chaves}},
  \bibinfo {author} {\bibfnamefont {J.}~\bibnamefont {Yu}}, \bibinfo {author}
  {\bibfnamefont {G.}~\bibnamefont {Arefe}}, \bibinfo {author} {\bibfnamefont
  {H.~M.}\ \bibnamefont {Hill}}, \bibinfo {author} {\bibfnamefont {A.~F.}\
  \bibnamefont {Rigosi}}, \bibinfo {author} {\bibfnamefont {T.~C.}\
  \bibnamefont {Berkelbach}}, \bibinfo {author} {\bibfnamefont
  {P.}~\bibnamefont {Nagler}}, \bibinfo {author} {\bibfnamefont
  {C.}~\bibnamefont {Sch{\"u}ller}}, \bibinfo {author} {\bibfnamefont
  {T.}~\bibnamefont {Korn}}, \bibinfo {author} {\bibfnamefont {C.}~\bibnamefont
  {Nuckolls}}, \bibinfo {author} {\bibfnamefont {J.}~\bibnamefont {Hone}},
  \bibinfo {author} {\bibfnamefont {L.~E.}\ \bibnamefont {Brus}}, \bibinfo
  {author} {\bibfnamefont {T.~F.}\ \bibnamefont {Heinz}}, \bibinfo {author}
  {\bibfnamefont {D.~R.}\ \bibnamefont {Reichman}},\ and\ \bibinfo {author}
  {\bibfnamefont {A.}~\bibnamefont {Chernikov}},\ }\bibfield  {title} {\bibinfo
  {title} {Coulomb engineering of the bandgap and excitons in two-dimensional
  materials},\ }\href {https://doi.org/10.1038/ncomms15251} {\bibfield
  {journal} {\bibinfo  {journal} {Nature Communications}\ }\textbf {\bibinfo
  {volume} {8}},\ \bibinfo {pages} {15251} (\bibinfo {year}
  {2017})}\BibitemShut {NoStop}%
\bibitem [{\citenamefont {Rytova}(1967)}]{rytova1967}%
  \BibitemOpen
  \bibfield  {author} {\bibinfo {author} {\bibfnamefont {H.~C.}\ \bibnamefont
  {Rytova}},\ }\bibfield  {title} {\bibinfo {title} {Coulomb interaction in
  thin semiconductor and semimetal films},\ }\href@noop {} {\bibfield
  {journal} {\bibinfo  {journal} {Moscow University Physics Bulletin}\ }\textbf
  {\bibinfo {volume} {3}},\ \bibinfo {pages} {18} (\bibinfo {year}
  {1967})}\BibitemShut {NoStop}%
\bibitem [{\citenamefont {Keldysh}(1979)}]{Keldysh1979}%
  \BibitemOpen
  \bibfield  {author} {\bibinfo {author} {\bibfnamefont {L.}~\bibnamefont
  {Keldysh}},\ }\bibfield  {title} {\bibinfo {title} {Coulomb interaction in
  thin semiconductor and semimetal films},\ }\href
  {http://www.jetpletters.ac.ru/ps/1458/article_22207.shtml} {\bibfield
  {journal} {\bibinfo  {journal} {JETP Lett.}\ }\textbf {\bibinfo {volume}
  {29}},\ \bibinfo {pages} {658} (\bibinfo {year} {1979})}\BibitemShut
  {NoStop}%
\bibitem [{\citenamefont {Qiu}\ \emph {et~al.}(2013)\citenamefont {Qiu},
  \citenamefont {da~Jornada},\ and\ \citenamefont {Louie}}]{Qiu2013}%
  \BibitemOpen
  \bibfield  {author} {\bibinfo {author} {\bibfnamefont {D.~Y.}\ \bibnamefont
  {Qiu}}, \bibinfo {author} {\bibfnamefont {F.~H.}\ \bibnamefont
  {da~Jornada}},\ and\ \bibinfo {author} {\bibfnamefont {S.~G.}\ \bibnamefont
  {Louie}},\ }\bibfield  {title} {\bibinfo {title} {Optical spectrum of
  ${\mathrm{mos}}_{2}$: Many-body effects and diversity of exciton states},\
  }\href {https://doi.org/10.1103/PhysRevLett.111.216805} {\bibfield  {journal}
  {\bibinfo  {journal} {Phys. Rev. Lett.}\ }\textbf {\bibinfo {volume} {111}},\
  \bibinfo {pages} {216805} (\bibinfo {year} {2013})}\BibitemShut {NoStop}%
\bibitem [{\citenamefont {Wang}\ \emph {et~al.}(2018)\citenamefont {Wang},
  \citenamefont {Chernikov}, \citenamefont {Glazov}, \citenamefont {Heinz},
  \citenamefont {Marie}, \citenamefont {Amand},\ and\ \citenamefont
  {Urbaszek}}]{Wang2018}%
  \BibitemOpen
  \bibfield  {author} {\bibinfo {author} {\bibfnamefont {G.}~\bibnamefont
  {Wang}}, \bibinfo {author} {\bibfnamefont {A.}~\bibnamefont {Chernikov}},
  \bibinfo {author} {\bibfnamefont {M.~M.}\ \bibnamefont {Glazov}}, \bibinfo
  {author} {\bibfnamefont {T.~F.}\ \bibnamefont {Heinz}}, \bibinfo {author}
  {\bibfnamefont {X.}~\bibnamefont {Marie}}, \bibinfo {author} {\bibfnamefont
  {T.}~\bibnamefont {Amand}},\ and\ \bibinfo {author} {\bibfnamefont
  {B.}~\bibnamefont {Urbaszek}},\ }\bibfield  {title} {\bibinfo {title}
  {Colloquium : Excitons in atomically thin transition metal dichalcogenides},\
  }\bibfield  {journal} {\bibinfo  {journal} {Reviews of Modern Physics}\
  }\textbf {\bibinfo {volume} {90}},\ \href
  {https://doi.org/10.1103/revmodphys.90.021001} {10.1103/revmodphys.90.021001}
  (\bibinfo {year} {2018})\BibitemShut {NoStop}%
\bibitem [{\citenamefont {Van~Tuan}\ \emph {et~al.}(2018)\citenamefont
  {Van~Tuan}, \citenamefont {Yang},\ and\ \citenamefont {Dery}}]{Dery2018}%
  \BibitemOpen
  \bibfield  {author} {\bibinfo {author} {\bibfnamefont {D.}~\bibnamefont
  {Van~Tuan}}, \bibinfo {author} {\bibfnamefont {M.}~\bibnamefont {Yang}},\
  and\ \bibinfo {author} {\bibfnamefont {H.}~\bibnamefont {Dery}},\ }\bibfield
  {title} {\bibinfo {title} {Coulomb interaction in monolayer transition-metal
  dichalcogenides},\ }\bibfield  {journal} {\bibinfo  {journal} {Physical
  Review B}\ }\textbf {\bibinfo {volume} {98}},\ \href
  {https://doi.org/10.1103/physrevb.98.125308} {10.1103/physrevb.98.125308}
  (\bibinfo {year} {2018})\BibitemShut {NoStop}%
\bibitem [{\citenamefont {Stier}\ \emph {et~al.}(2018)\citenamefont {Stier},
  \citenamefont {Wilson}, \citenamefont {Velizhanin}, \citenamefont {Kono},
  \citenamefont {Xu},\ and\ \citenamefont {Crooker}}]{stier2018}%
  \BibitemOpen
  \bibfield  {author} {\bibinfo {author} {\bibfnamefont {A.~V.}\ \bibnamefont
  {Stier}}, \bibinfo {author} {\bibfnamefont {N.~P.}\ \bibnamefont {Wilson}},
  \bibinfo {author} {\bibfnamefont {K.~A.}\ \bibnamefont {Velizhanin}},
  \bibinfo {author} {\bibfnamefont {J.}~\bibnamefont {Kono}}, \bibinfo {author}
  {\bibfnamefont {X.}~\bibnamefont {Xu}},\ and\ \bibinfo {author}
  {\bibfnamefont {S.~A.}\ \bibnamefont {Crooker}},\ }\bibfield  {title}
  {\bibinfo {title} {Magnetooptics of exciton rydberg states in a monolayer
  semiconductor},\ }\href {https://doi.org/10.1103/PhysRevLett.120.057405}
  {\bibfield  {journal} {\bibinfo  {journal} {Phys. Rev. Lett.}\ }\textbf
  {\bibinfo {volume} {120}},\ \bibinfo {pages} {057405} (\bibinfo {year}
  {2018})}\BibitemShut {NoStop}%
\bibitem [{\citenamefont {Molas}\ \emph {et~al.}(2019)\citenamefont {Molas},
  \citenamefont {Slobodeniuk}, \citenamefont {Nogajewski}, \citenamefont
  {Bartos}, \citenamefont {Bala}, \citenamefont {Babiński}, \citenamefont
  {Watanabe}, \citenamefont {Taniguchi}, \citenamefont {Faugeras},\ and\
  \citenamefont {Potemski}}]{Molas2019}%
  \BibitemOpen
  \bibfield  {author} {\bibinfo {author} {\bibfnamefont {M.~R.}\ \bibnamefont
  {Molas}}, \bibinfo {author} {\bibfnamefont {A.}~\bibnamefont {Slobodeniuk}},
  \bibinfo {author} {\bibfnamefont {K.}~\bibnamefont {Nogajewski}}, \bibinfo
  {author} {\bibfnamefont {M.}~\bibnamefont {Bartos}}, \bibinfo {author}
  {\bibfnamefont {L.}~\bibnamefont {Bala}}, \bibinfo {author} {\bibfnamefont
  {A.}~\bibnamefont {Babiński}}, \bibinfo {author} {\bibfnamefont
  {K.}~\bibnamefont {Watanabe}}, \bibinfo {author} {\bibfnamefont
  {T.}~\bibnamefont {Taniguchi}}, \bibinfo {author} {\bibfnamefont
  {C.}~\bibnamefont {Faugeras}},\ and\ \bibinfo {author} {\bibfnamefont
  {M.}~\bibnamefont {Potemski}},\ }\bibfield  {title} {\bibinfo {title} {Energy
  spectrum of two-dimensional excitons in a nonuniform dielectric medium},\
  }\bibfield  {journal} {\bibinfo  {journal} {Physical Review Letters}\
  }\textbf {\bibinfo {volume} {123}},\ \href
  {https://doi.org/10.1103/physrevlett.123.136801}
  {10.1103/physrevlett.123.136801} (\bibinfo {year} {2019})\BibitemShut
  {NoStop}%
\bibitem [{\citenamefont {Cadiz}\ \emph {et~al.}(2017)\citenamefont {Cadiz},
  \citenamefont {Courtade}, \citenamefont {Robert}, \citenamefont {Wang},
  \citenamefont {Shen}, \citenamefont {Cai}, \citenamefont {Taniguchi},
  \citenamefont {Watanabe}, \citenamefont {Carrere}, \citenamefont {Lagarde},
  \citenamefont {Manca}, \citenamefont {Amand}, \citenamefont {Renucci},
  \citenamefont {Tongay}, \citenamefont {Marie},\ and\ \citenamefont
  {Urbaszek}}]{Cadiz2017}%
  \BibitemOpen
  \bibfield  {author} {\bibinfo {author} {\bibfnamefont {F.}~\bibnamefont
  {Cadiz}}, \bibinfo {author} {\bibfnamefont {E.}~\bibnamefont {Courtade}},
  \bibinfo {author} {\bibfnamefont {C.}~\bibnamefont {Robert}}, \bibinfo
  {author} {\bibfnamefont {G.}~\bibnamefont {Wang}}, \bibinfo {author}
  {\bibfnamefont {Y.}~\bibnamefont {Shen}}, \bibinfo {author} {\bibfnamefont
  {H.}~\bibnamefont {Cai}}, \bibinfo {author} {\bibfnamefont {T.}~\bibnamefont
  {Taniguchi}}, \bibinfo {author} {\bibfnamefont {K.}~\bibnamefont {Watanabe}},
  \bibinfo {author} {\bibfnamefont {H.}~\bibnamefont {Carrere}}, \bibinfo
  {author} {\bibfnamefont {D.}~\bibnamefont {Lagarde}}, \bibinfo {author}
  {\bibfnamefont {M.}~\bibnamefont {Manca}}, \bibinfo {author} {\bibfnamefont
  {T.}~\bibnamefont {Amand}}, \bibinfo {author} {\bibfnamefont
  {P.}~\bibnamefont {Renucci}}, \bibinfo {author} {\bibfnamefont
  {S.}~\bibnamefont {Tongay}}, \bibinfo {author} {\bibfnamefont
  {X.}~\bibnamefont {Marie}},\ and\ \bibinfo {author} {\bibfnamefont
  {B.}~\bibnamefont {Urbaszek}},\ }\bibfield  {title} {\bibinfo {title}
  {Excitonic linewidth approaching the homogeneous limit in
  ${\mathrm{mos}}_{2}$-based van der waals heterostructures},\ }\href
  {https://doi.org/10.1103/PhysRevX.7.021026} {\bibfield  {journal} {\bibinfo
  {journal} {Phys. Rev. X}\ }\textbf {\bibinfo {volume} {7}},\ \bibinfo {pages}
  {021026} (\bibinfo {year} {2017})}\BibitemShut {NoStop}%
\bibitem [{\citenamefont {Hsu}\ \emph {et~al.}(2022)\citenamefont {Hsu},
  \citenamefont {Quan}, \citenamefont {Pan}, \citenamefont {Chen},
  \citenamefont {Chou}, \citenamefont {Chang}, \citenamefont {MacDonald},
  \citenamefont {Li}, \citenamefont {Lin},\ and\ \citenamefont
  {Shih}}]{Hsu2022}%
  \BibitemOpen
  \bibfield  {author} {\bibinfo {author} {\bibfnamefont {W.-T.}\ \bibnamefont
  {Hsu}}, \bibinfo {author} {\bibfnamefont {J.}~\bibnamefont {Quan}}, \bibinfo
  {author} {\bibfnamefont {C.-R.}\ \bibnamefont {Pan}}, \bibinfo {author}
  {\bibfnamefont {P.-J.}\ \bibnamefont {Chen}}, \bibinfo {author}
  {\bibfnamefont {M.-Y.}\ \bibnamefont {Chou}}, \bibinfo {author}
  {\bibfnamefont {W.-H.}\ \bibnamefont {Chang}}, \bibinfo {author}
  {\bibfnamefont {A.~H.}\ \bibnamefont {MacDonald}}, \bibinfo {author}
  {\bibfnamefont {X.}~\bibnamefont {Li}}, \bibinfo {author} {\bibfnamefont
  {J.-F.}\ \bibnamefont {Lin}},\ and\ \bibinfo {author} {\bibfnamefont {C.-K.}\
  \bibnamefont {Shih}},\ }\bibfield  {title} {\bibinfo {title} {Quantitative
  determination of interlayer electronic coupling at various critical points in
  bilayer ${\mathrm{mos}}_{2}$},\ }\bibfield  {journal} {\bibinfo  {journal}
  {Physical Review B}\ }\textbf {\bibinfo {volume} {106}},\ \href
  {https://doi.org/10.1103/physrevb.106.125302} {10.1103/physrevb.106.125302}
  (\bibinfo {year} {2022})\BibitemShut {NoStop}%
\bibitem [{\citenamefont {Xia}\ \emph {et~al.}(2020)\citenamefont {Xia},
  \citenamefont {Yan}, \citenamefont {Wang}, \citenamefont {He}, \citenamefont
  {Gong}, \citenamefont {Chen}, \citenamefont {Sum}, \citenamefont {Liu},
  \citenamefont {Ajayan},\ and\ \citenamefont {Shen}}]{Xia2020}%
  \BibitemOpen
  \bibfield  {author} {\bibinfo {author} {\bibfnamefont {J.}~\bibnamefont
  {Xia}}, \bibinfo {author} {\bibfnamefont {J.}~\bibnamefont {Yan}}, \bibinfo
  {author} {\bibfnamefont {Z.}~\bibnamefont {Wang}}, \bibinfo {author}
  {\bibfnamefont {Y.}~\bibnamefont {He}}, \bibinfo {author} {\bibfnamefont
  {Y.}~\bibnamefont {Gong}}, \bibinfo {author} {\bibfnamefont {W.}~\bibnamefont
  {Chen}}, \bibinfo {author} {\bibfnamefont {T.~C.}\ \bibnamefont {Sum}},
  \bibinfo {author} {\bibfnamefont {Z.}~\bibnamefont {Liu}}, \bibinfo {author}
  {\bibfnamefont {P.~M.}\ \bibnamefont {Ajayan}},\ and\ \bibinfo {author}
  {\bibfnamefont {Z.}~\bibnamefont {Shen}},\ }\bibfield  {title} {\bibinfo
  {title} {Strong coupling and pressure engineering in wse2–mose2
  heterobilayers},\ }\href {https://doi.org/10.1038/s41567-020-1005-7}
  {\bibfield  {journal} {\bibinfo  {journal} {Nature Physics}\ }\textbf
  {\bibinfo {volume} {17}},\ \bibinfo {pages} {92–98} (\bibinfo {year}
  {2020})}\BibitemShut {NoStop}%
\bibitem [{\citenamefont {Zhu}\ \emph {et~al.}(2022)\citenamefont {Zhu},
  \citenamefont {Zhang}, \citenamefont {Zhang}, \citenamefont {Liu},
  \citenamefont {Zhang}, \citenamefont {Zhang}, \citenamefont {Li},
  \citenamefont {Cheng},\ and\ \citenamefont {Huang}}]{Zhu2022}%
  \BibitemOpen
  \bibfield  {author} {\bibinfo {author} {\bibfnamefont {M.}~\bibnamefont
  {Zhu}}, \bibinfo {author} {\bibfnamefont {Z.}~\bibnamefont {Zhang}}, \bibinfo
  {author} {\bibfnamefont {T.}~\bibnamefont {Zhang}}, \bibinfo {author}
  {\bibfnamefont {D.}~\bibnamefont {Liu}}, \bibinfo {author} {\bibfnamefont
  {H.}~\bibnamefont {Zhang}}, \bibinfo {author} {\bibfnamefont
  {Z.}~\bibnamefont {Zhang}}, \bibinfo {author} {\bibfnamefont
  {Z.}~\bibnamefont {Li}}, \bibinfo {author} {\bibfnamefont {Y.}~\bibnamefont
  {Cheng}},\ and\ \bibinfo {author} {\bibfnamefont {W.}~\bibnamefont {Huang}},\
  }\bibfield  {title} {\bibinfo {title} {Exchange between interlayer and
  intralayer exciton in wse2/ws2 heterostructure by interlayer coupling
  engineering},\ }\href {https://doi.org/10.1021/acs.nanolett.2c01353}
  {\bibfield  {journal} {\bibinfo  {journal} {Nano Letters}\ }\textbf {\bibinfo
  {volume} {22}},\ \bibinfo {pages} {4528–4534} (\bibinfo {year}
  {2022})}\BibitemShut {NoStop}%
\bibitem [{\citenamefont {Yankowitz}\ \emph {et~al.}(2019)\citenamefont
  {Yankowitz}, \citenamefont {Chen}, \citenamefont {Polshyn}, \citenamefont
  {Zhang}, \citenamefont {Watanabe}, \citenamefont {Taniguchi}, \citenamefont
  {Graf}, \citenamefont {Young},\ and\ \citenamefont {Dean}}]{Yankowitz2019}%
  \BibitemOpen
  \bibfield  {author} {\bibinfo {author} {\bibfnamefont {M.}~\bibnamefont
  {Yankowitz}}, \bibinfo {author} {\bibfnamefont {S.}~\bibnamefont {Chen}},
  \bibinfo {author} {\bibfnamefont {H.}~\bibnamefont {Polshyn}}, \bibinfo
  {author} {\bibfnamefont {Y.}~\bibnamefont {Zhang}}, \bibinfo {author}
  {\bibfnamefont {K.}~\bibnamefont {Watanabe}}, \bibinfo {author}
  {\bibfnamefont {T.}~\bibnamefont {Taniguchi}}, \bibinfo {author}
  {\bibfnamefont {D.}~\bibnamefont {Graf}}, \bibinfo {author} {\bibfnamefont
  {A.~F.}\ \bibnamefont {Young}},\ and\ \bibinfo {author} {\bibfnamefont
  {C.~R.}\ \bibnamefont {Dean}},\ }\bibfield  {title} {\bibinfo {title} {Tuning
  superconductivity in twisted bilayer graphene},\ }\href
  {https://doi.org/10.1126/science.aav1910} {\bibfield  {journal} {\bibinfo
  {journal} {Science}\ }\textbf {\bibinfo {volume} {363}},\ \bibinfo {pages}
  {1059–1064} (\bibinfo {year} {2019})}\BibitemShut {NoStop}%
\bibitem [{\citenamefont {Pawbake}\ \emph {et~al.}(2023)\citenamefont
  {Pawbake}, \citenamefont {Pelini}, \citenamefont {Mohelsky}, \citenamefont
  {Jana}, \citenamefont {Breslavetz}, \citenamefont {Cho}, \citenamefont
  {Orlita}, \citenamefont {Potemski}, \citenamefont {Measson}, \citenamefont
  {Wilson} \emph {et~al.}}]{Pawbake2023}%
  \BibitemOpen
  \bibfield  {author} {\bibinfo {author} {\bibfnamefont {A.}~\bibnamefont
  {Pawbake}}, \bibinfo {author} {\bibfnamefont {T.}~\bibnamefont {Pelini}},
  \bibinfo {author} {\bibfnamefont {I.}~\bibnamefont {Mohelsky}}, \bibinfo
  {author} {\bibfnamefont {D.}~\bibnamefont {Jana}}, \bibinfo {author}
  {\bibfnamefont {I.}~\bibnamefont {Breslavetz}}, \bibinfo {author}
  {\bibfnamefont {C.-W.}\ \bibnamefont {Cho}}, \bibinfo {author} {\bibfnamefont
  {M.}~\bibnamefont {Orlita}}, \bibinfo {author} {\bibfnamefont
  {M.}~\bibnamefont {Potemski}}, \bibinfo {author} {\bibfnamefont {M.-A.}\
  \bibnamefont {Measson}}, \bibinfo {author} {\bibfnamefont {N.~P.}\
  \bibnamefont {Wilson}}, \emph {et~al.},\ }\bibfield  {title} {\bibinfo
  {title} {Magneto-optical sensing of the pressure driven magnetic ground
  states in bulk crsbr},\ }\href@noop {} {\bibfield  {journal} {\bibinfo
  {journal} {Nano Letters}\ }\textbf {\bibinfo {volume} {23}},\ \bibinfo
  {pages} {9587} (\bibinfo {year} {2023})}\BibitemShut {NoStop}%
\bibitem [{\citenamefont {Pawbake}\ \emph {et~al.}(2022)\citenamefont
  {Pawbake}, \citenamefont {Pelini}, \citenamefont {Delhomme}, \citenamefont
  {Romanin}, \citenamefont {Vaclavkova}, \citenamefont {Martinez},
  \citenamefont {Calandra}, \citenamefont {Measson}, \citenamefont {Veis},
  \citenamefont {Potemski}, \citenamefont {Orlita},\ and\ \citenamefont
  {Faugeras}}]{Pawbake2022}%
  \BibitemOpen
  \bibfield  {author} {\bibinfo {author} {\bibfnamefont {A.}~\bibnamefont
  {Pawbake}}, \bibinfo {author} {\bibfnamefont {T.}~\bibnamefont {Pelini}},
  \bibinfo {author} {\bibfnamefont {A.}~\bibnamefont {Delhomme}}, \bibinfo
  {author} {\bibfnamefont {D.}~\bibnamefont {Romanin}}, \bibinfo {author}
  {\bibfnamefont {D.}~\bibnamefont {Vaclavkova}}, \bibinfo {author}
  {\bibfnamefont {G.}~\bibnamefont {Martinez}}, \bibinfo {author}
  {\bibfnamefont {M.}~\bibnamefont {Calandra}}, \bibinfo {author}
  {\bibfnamefont {M.-A.}\ \bibnamefont {Measson}}, \bibinfo {author}
  {\bibfnamefont {M.}~\bibnamefont {Veis}}, \bibinfo {author} {\bibfnamefont
  {M.}~\bibnamefont {Potemski}}, \bibinfo {author} {\bibfnamefont
  {M.}~\bibnamefont {Orlita}},\ and\ \bibinfo {author} {\bibfnamefont
  {C.}~\bibnamefont {Faugeras}},\ }\bibfield  {title} {\bibinfo {title}
  {High-pressure tuning of magnon-polarons in the layered antiferromagnet
  feps3},\ }\href {https://doi.org/10.1021/acsnano.2c04286} {\bibfield
  {journal} {\bibinfo  {journal} {ACS Nano}\ }\textbf {\bibinfo {volume}
  {16}},\ \bibinfo {pages} {12656–12665} (\bibinfo {year}
  {2022})}\BibitemShut {NoStop}%
\bibitem [{\citenamefont {Castellanos-Gomez}\ \emph {et~al.}(2014)\citenamefont
  {Castellanos-Gomez}, \citenamefont {Buscema}, \citenamefont {Molenaar},
  \citenamefont {Singh}, \citenamefont {Janssen}, \citenamefont {van~der
  Zant},\ and\ \citenamefont {Steele}}]{gomez2014}%
  \BibitemOpen
  \bibfield  {author} {\bibinfo {author} {\bibfnamefont {A.}~\bibnamefont
  {Castellanos-Gomez}}, \bibinfo {author} {\bibfnamefont {M.}~\bibnamefont
  {Buscema}}, \bibinfo {author} {\bibfnamefont {R.}~\bibnamefont {Molenaar}},
  \bibinfo {author} {\bibfnamefont {V.}~\bibnamefont {Singh}}, \bibinfo
  {author} {\bibfnamefont {L.}~\bibnamefont {Janssen}}, \bibinfo {author}
  {\bibfnamefont {H.~S.~J.}\ \bibnamefont {van~der Zant}},\ and\ \bibinfo
  {author} {\bibfnamefont {G.~A.}\ \bibnamefont {Steele}},\ }\bibfield  {title}
  {\bibinfo {title} {Deterministic transfer of two-dimensional materials by
  all-dry viscoelastic stamping},\ }\href
  {https://doi.org/10.1088/2053-1583/1/1/011002} {\bibfield  {journal}
  {\bibinfo  {journal} {2D Materials}\ }\textbf {\bibinfo {volume} {1}},\
  \bibinfo {pages} {011002} (\bibinfo {year} {2014})}\BibitemShut {NoStop}%
\bibitem [{\citenamefont {Goossens}\ \emph {et~al.}(2012)\citenamefont
  {Goossens}, \citenamefont {Calado}, \citenamefont {Barreiro}, \citenamefont
  {Watanabe}, \citenamefont {Taniguchi},\ and\ \citenamefont
  {Vandersypen}}]{Goossens2012}%
  \BibitemOpen
  \bibfield  {author} {\bibinfo {author} {\bibfnamefont {A.~M.}\ \bibnamefont
  {Goossens}}, \bibinfo {author} {\bibfnamefont {V.~E.}\ \bibnamefont
  {Calado}}, \bibinfo {author} {\bibfnamefont {A.}~\bibnamefont {Barreiro}},
  \bibinfo {author} {\bibfnamefont {K.}~\bibnamefont {Watanabe}}, \bibinfo
  {author} {\bibfnamefont {T.}~\bibnamefont {Taniguchi}},\ and\ \bibinfo
  {author} {\bibfnamefont {L.~M.~K.}\ \bibnamefont {Vandersypen}},\ }\bibfield
  {title} {\bibinfo {title} {Mechanical cleaning of graphene},\ }\bibfield
  {journal} {\bibinfo  {journal} {Applied Physics Letters}\ }\textbf {\bibinfo
  {volume} {100}},\ \href {https://doi.org/10.1063/1.3685504}
  {10.1063/1.3685504} (\bibinfo {year} {2012})\BibitemShut {NoStop}%
\bibitem [{\citenamefont {Kim}\ \emph {et~al.}(2019)\citenamefont {Kim},
  \citenamefont {Herlinger}, \citenamefont {Taniguchi}, \citenamefont
  {Watanabe},\ and\ \citenamefont {Smet}}]{Kim2019}%
  \BibitemOpen
  \bibfield  {author} {\bibinfo {author} {\bibfnamefont {Y.}~\bibnamefont
  {Kim}}, \bibinfo {author} {\bibfnamefont {P.}~\bibnamefont {Herlinger}},
  \bibinfo {author} {\bibfnamefont {T.}~\bibnamefont {Taniguchi}}, \bibinfo
  {author} {\bibfnamefont {K.}~\bibnamefont {Watanabe}},\ and\ \bibinfo
  {author} {\bibfnamefont {J.~H.}\ \bibnamefont {Smet}},\ }\bibfield  {title}
  {\bibinfo {title} {Reliable postprocessing improvement of van der waals
  heterostructures},\ }\href {https://doi.org/10.1021/acsnano.9b06992}
  {\bibfield  {journal} {\bibinfo  {journal} {ACS Nano}\ }\textbf {\bibinfo
  {volume} {13}},\ \bibinfo {pages} {14182–14190} (\bibinfo {year}
  {2019})}\BibitemShut {NoStop}%
\bibitem [{\citenamefont {Breslavetz}\ \emph {et~al.}(2021)\citenamefont
  {Breslavetz}, \citenamefont {Delhomme}, \citenamefont {Pelini}, \citenamefont
  {Pawbake}, \citenamefont {Vaclavkova}, \citenamefont {Orlita}, \citenamefont
  {Potemski}, \citenamefont {Measson},\ and\ \citenamefont
  {Faugeras}}]{Breslavetz2021}%
  \BibitemOpen
  \bibfield  {author} {\bibinfo {author} {\bibfnamefont {I.}~\bibnamefont
  {Breslavetz}}, \bibinfo {author} {\bibfnamefont {A.}~\bibnamefont
  {Delhomme}}, \bibinfo {author} {\bibfnamefont {T.}~\bibnamefont {Pelini}},
  \bibinfo {author} {\bibfnamefont {A.}~\bibnamefont {Pawbake}}, \bibinfo
  {author} {\bibfnamefont {D.}~\bibnamefont {Vaclavkova}}, \bibinfo {author}
  {\bibfnamefont {M.}~\bibnamefont {Orlita}}, \bibinfo {author} {\bibfnamefont
  {M.}~\bibnamefont {Potemski}}, \bibinfo {author} {\bibfnamefont {M.-A.}\
  \bibnamefont {Measson}},\ and\ \bibinfo {author} {\bibfnamefont
  {C.}~\bibnamefont {Faugeras}},\ }\bibfield  {title} {\bibinfo {title}
  {Spatially resolved optical spectroscopy in extreme environment of low
  temperature, high magnetic fields and high pressure},\ }\bibfield  {journal}
  {\bibinfo  {journal} {Review of Scientific Instruments}\ }\textbf {\bibinfo
  {volume} {92}},\ \href {https://doi.org/10.1063/5.0070934}
  {10.1063/5.0070934} (\bibinfo {year} {2021})\BibitemShut {NoStop}%
\bibitem [{\citenamefont {Latini}\ \emph {et~al.}(2015)\citenamefont {Latini},
  \citenamefont {Olsen},\ and\ \citenamefont {Thygesen}}]{Latini2015}%
  \BibitemOpen
  \bibfield  {author} {\bibinfo {author} {\bibfnamefont {S.}~\bibnamefont
  {Latini}}, \bibinfo {author} {\bibfnamefont {T.}~\bibnamefont {Olsen}},\ and\
  \bibinfo {author} {\bibfnamefont {K.~S.}\ \bibnamefont {Thygesen}},\
  }\bibfield  {title} {\bibinfo {title} {Excitons in van der waals
  heterostructures: The important role of dielectric screening},\ }\bibfield
  {journal} {\bibinfo  {journal} {Physical Review B}\ }\textbf {\bibinfo
  {volume} {92}},\ \href {https://doi.org/10.1103/physrevb.92.245123}
  {10.1103/physrevb.92.245123} (\bibinfo {year} {2015})\BibitemShut {NoStop}%
\bibitem [{\citenamefont {Florian}\ \emph {et~al.}(2018)\citenamefont
  {Florian}, \citenamefont {Hartmann}, \citenamefont {Steinhoff}, \citenamefont
  {Klein}, \citenamefont {Holleitner}, \citenamefont {Finley}, \citenamefont
  {Wehling}, \citenamefont {Kaniber},\ and\ \citenamefont
  {Gies}}]{Florian2018}%
  \BibitemOpen
  \bibfield  {author} {\bibinfo {author} {\bibfnamefont {M.}~\bibnamefont
  {Florian}}, \bibinfo {author} {\bibfnamefont {M.}~\bibnamefont {Hartmann}},
  \bibinfo {author} {\bibfnamefont {A.}~\bibnamefont {Steinhoff}}, \bibinfo
  {author} {\bibfnamefont {J.}~\bibnamefont {Klein}}, \bibinfo {author}
  {\bibfnamefont {A.~W.}\ \bibnamefont {Holleitner}}, \bibinfo {author}
  {\bibfnamefont {J.~J.}\ \bibnamefont {Finley}}, \bibinfo {author}
  {\bibfnamefont {T.~O.}\ \bibnamefont {Wehling}}, \bibinfo {author}
  {\bibfnamefont {M.}~\bibnamefont {Kaniber}},\ and\ \bibinfo {author}
  {\bibfnamefont {C.}~\bibnamefont {Gies}},\ }\bibfield  {title} {\bibinfo
  {title} {The dielectric impact of layer distances on exciton and trion
  binding energies in van der waals heterostructures},\ }\href
  {https://doi.org/10.1021/acs.nanolett.8b00840} {\bibfield  {journal}
  {\bibinfo  {journal} {Nano Letters}\ }\textbf {\bibinfo {volume} {18}},\
  \bibinfo {pages} {2725–2732} (\bibinfo {year} {2018})}\BibitemShut
  {NoStop}%
\bibitem [{\citenamefont {R\"{o}sner}\ \emph {et~al.}(2015)\citenamefont
  {R\"{o}sner}, \citenamefont {Şaşıoğlu}, \citenamefont {Friedrich},
  \citenamefont {Bl\"{u}gel},\ and\ \citenamefont {Wehling}}]{rosner2015}%
  \BibitemOpen
  \bibfield  {author} {\bibinfo {author} {\bibfnamefont {M.}~\bibnamefont
  {R\"{o}sner}}, \bibinfo {author} {\bibfnamefont {E.}~\bibnamefont
  {Şaşıoğlu}}, \bibinfo {author} {\bibfnamefont {C.}~\bibnamefont
  {Friedrich}}, \bibinfo {author} {\bibfnamefont {S.}~\bibnamefont
  {Bl\"{u}gel}},\ and\ \bibinfo {author} {\bibfnamefont {T.~O.}\ \bibnamefont
  {Wehling}},\ }\bibfield  {title} {\bibinfo {title} {Wannier function approach
  to realistic coulomb interactions in layered materials and
  heterostructures},\ }\bibfield  {journal} {\bibinfo  {journal} {Physical
  Review B}\ }\textbf {\bibinfo {volume} {92}},\ \href
  {https://doi.org/10.1103/physrevb.92.085102} {10.1103/physrevb.92.085102}
  (\bibinfo {year} {2015})\BibitemShut {NoStop}%
\bibitem [{\citenamefont {Kresse}\ and\ \citenamefont
  {Hafner}(1993)}]{Kresse1993}%
  \BibitemOpen
  \bibfield  {author} {\bibinfo {author} {\bibfnamefont {G.}~\bibnamefont
  {Kresse}}\ and\ \bibinfo {author} {\bibfnamefont {J.}~\bibnamefont
  {Hafner}},\ }\bibfield  {title} {\bibinfo {title} {Ab initiomolecular
  dynamics for liquid metals},\ }\href
  {https://doi.org/10.1103/physrevb.47.558} {\bibfield  {journal} {\bibinfo
  {journal} {Physical Review B}\ }\textbf {\bibinfo {volume} {47}},\ \bibinfo
  {pages} {558–561} (\bibinfo {year} {1993})}\BibitemShut {NoStop}%
\bibitem [{\citenamefont {Kresse}\ and\ \citenamefont
  {Furthm\"{u}ller}(1996)}]{Kresse1996}%
  \BibitemOpen
  \bibfield  {author} {\bibinfo {author} {\bibfnamefont {G.}~\bibnamefont
  {Kresse}}\ and\ \bibinfo {author} {\bibfnamefont {J.}~\bibnamefont
  {Furthm\"{u}ller}},\ }\bibfield  {title} {\bibinfo {title} {Efficient
  iterative schemes forab initiototal-energy calculations using a plane-wave
  basis set},\ }\href {https://doi.org/10.1103/physrevb.54.11169} {\bibfield
  {journal} {\bibinfo  {journal} {Physical Review B}\ }\textbf {\bibinfo
  {volume} {54}},\ \bibinfo {pages} {11169–11186} (\bibinfo {year}
  {1996})}\BibitemShut {NoStop}%
\bibitem [{\citenamefont {Perdew}\ \emph {et~al.}(1996)\citenamefont {Perdew},
  \citenamefont {Burke},\ and\ \citenamefont {Ernzerhof}}]{Perdew1996}%
  \BibitemOpen
  \bibfield  {author} {\bibinfo {author} {\bibfnamefont {J.~P.}\ \bibnamefont
  {Perdew}}, \bibinfo {author} {\bibfnamefont {K.}~\bibnamefont {Burke}},\ and\
  \bibinfo {author} {\bibfnamefont {M.}~\bibnamefont {Ernzerhof}},\ }\bibfield
  {title} {\bibinfo {title} {Generalized gradient approximation made simple},\
  }\href {https://doi.org/10.1103/physrevlett.77.3865} {\bibfield  {journal}
  {\bibinfo  {journal} {Physical Review Letters}\ }\textbf {\bibinfo {volume}
  {77}},\ \bibinfo {pages} {3865–3868} (\bibinfo {year} {1996})}\BibitemShut
  {NoStop}%
\bibitem [{\citenamefont {Blochl}(1994)}]{Blochl1994}%
  \BibitemOpen
  \bibfield  {author} {\bibinfo {author} {\bibfnamefont {P.~E.}\ \bibnamefont
  {Blochl}},\ }\bibfield  {title} {\bibinfo {title} {Projector augmented-wave
  method},\ }\href {https://doi.org/10.1103/physrevb.50.17953} {\bibfield
  {journal} {\bibinfo  {journal} {Physical Review B}\ }\textbf {\bibinfo
  {volume} {50}},\ \bibinfo {pages} {17953–17979} (\bibinfo {year}
  {1994})}\BibitemShut {NoStop}%
\bibitem [{\citenamefont {Kresse}\ and\ \citenamefont
  {Joubert}(1999)}]{Kresse1999}%
  \BibitemOpen
  \bibfield  {author} {\bibinfo {author} {\bibfnamefont {G.}~\bibnamefont
  {Kresse}}\ and\ \bibinfo {author} {\bibfnamefont {D.}~\bibnamefont
  {Joubert}},\ }\bibfield  {title} {\bibinfo {title} {From ultrasoft
  pseudopotentials to the projector augmented-wave method},\ }\href
  {https://doi.org/10.1103/physrevb.59.1758} {\bibfield  {journal} {\bibinfo
  {journal} {Physical Review B}\ }\textbf {\bibinfo {volume} {59}},\ \bibinfo
  {pages} {1758–1775} (\bibinfo {year} {1999})}\BibitemShut {NoStop}%
\bibitem [{\citenamefont {Klimes}\ \emph {et~al.}(2009)\citenamefont {Klimes},
  \citenamefont {Bowler},\ and\ \citenamefont {Michaelides}}]{Klime2009}%
  \BibitemOpen
  \bibfield  {author} {\bibinfo {author} {\bibfnamefont {J.}~\bibnamefont
  {Klimes}}, \bibinfo {author} {\bibfnamefont {D.~R.}\ \bibnamefont {Bowler}},\
  and\ \bibinfo {author} {\bibfnamefont {A.}~\bibnamefont {Michaelides}},\
  }\bibfield  {title} {\bibinfo {title} {Chemical accuracy for the van der
  waals density functional},\ }\href
  {https://doi.org/10.1088/0953-8984/22/2/022201} {\bibfield  {journal}
  {\bibinfo  {journal} {Journal of Physics: Condensed Matter}\ }\textbf
  {\bibinfo {volume} {22}},\ \bibinfo {pages} {022201} (\bibinfo {year}
  {2009})}\BibitemShut {NoStop}%
\bibitem [{\citenamefont {Wang}\ \emph {et~al.}(2021)\citenamefont {Wang},
  \citenamefont {Xu}, \citenamefont {Liu}, \citenamefont {Tang},\ and\
  \citenamefont {Geng}}]{Wang2021}%
  \BibitemOpen
  \bibfield  {author} {\bibinfo {author} {\bibfnamefont {V.}~\bibnamefont
  {Wang}}, \bibinfo {author} {\bibfnamefont {N.}~\bibnamefont {Xu}}, \bibinfo
  {author} {\bibfnamefont {J.-C.}\ \bibnamefont {Liu}}, \bibinfo {author}
  {\bibfnamefont {G.}~\bibnamefont {Tang}},\ and\ \bibinfo {author}
  {\bibfnamefont {W.-T.}\ \bibnamefont {Geng}},\ }\bibfield  {title} {\bibinfo
  {title} {Vaspkit: A user-friendly interface facilitating high-throughput
  computing and analysis using vasp code},\ }\href
  {https://doi.org/10.1016/j.cpc.2021.108033} {\bibfield  {journal} {\bibinfo
  {journal} {Computer Physics Communications}\ }\textbf {\bibinfo {volume}
  {267}},\ \bibinfo {pages} {108033} (\bibinfo {year} {2021})}\BibitemShut
  {NoStop}%
\bibitem [{\citenamefont {Rybkovskiy}\ \emph {et~al.}(2017)\citenamefont
  {Rybkovskiy}, \citenamefont {Gerber},\ and\ \citenamefont
  {Durnev}}]{Rybkovskiy2017}%
  \BibitemOpen
  \bibfield  {author} {\bibinfo {author} {\bibfnamefont {D.~V.}\ \bibnamefont
  {Rybkovskiy}}, \bibinfo {author} {\bibfnamefont {I.~C.}\ \bibnamefont
  {Gerber}},\ and\ \bibinfo {author} {\bibfnamefont {M.~V.}\ \bibnamefont
  {Durnev}},\ }\bibfield  {title} {\bibinfo {title} {Atomically inspired k.p
  approach and valley zeeman effect in transition metal dichalcogenide
  monolayers},\ }\bibfield  {journal} {\bibinfo  {journal} {Physical Review B}\
  }\textbf {\bibinfo {volume} {95}},\ \href
  {https://doi.org/10.1103/physrevb.95.155406} {10.1103/physrevb.95.155406}
  (\bibinfo {year} {2017})\BibitemShut {NoStop}%
\bibitem [{\citenamefont {Popert}\ \emph {et~al.}(2022)\citenamefont {Popert},
  \citenamefont {Shimazaki}, \citenamefont {Kroner}, \citenamefont {Watanabe},
  \citenamefont {Taniguchi}, \citenamefont {Imamoğlu},\ and\ \citenamefont
  {Smoleński}}]{Popert2022}%
  \BibitemOpen
  \bibfield  {author} {\bibinfo {author} {\bibfnamefont {A.}~\bibnamefont
  {Popert}}, \bibinfo {author} {\bibfnamefont {Y.}~\bibnamefont {Shimazaki}},
  \bibinfo {author} {\bibfnamefont {M.}~\bibnamefont {Kroner}}, \bibinfo
  {author} {\bibfnamefont {K.}~\bibnamefont {Watanabe}}, \bibinfo {author}
  {\bibfnamefont {T.}~\bibnamefont {Taniguchi}}, \bibinfo {author}
  {\bibfnamefont {A.}~\bibnamefont {Imamoğlu}},\ and\ \bibinfo {author}
  {\bibfnamefont {T.}~\bibnamefont {Smoleński}},\ }\bibfield  {title}
  {\bibinfo {title} {Optical sensing of fractional quantum hall effect in
  graphene},\ }\href {https://doi.org/10.1021/acs.nanolett.2c02000} {\bibfield
  {journal} {\bibinfo  {journal} {Nano Letters}\ }\textbf {\bibinfo {volume}
  {22}},\ \bibinfo {pages} {7363–7369} (\bibinfo {year} {2022})}\BibitemShut
  {NoStop}%
\bibitem [{\citenamefont {He}\ \emph {et~al.}(2024)\citenamefont {He},
  \citenamefont {Cai}, \citenamefont {Zheng}, \citenamefont {Seewald},
  \citenamefont {Taniguchi}, \citenamefont {Watanabe}, \citenamefont {Yan},
  \citenamefont {Yankowitz}, \citenamefont {Pasupathy}, \citenamefont {Yao},\
  and\ \citenamefont {Xu}}]{He2024}%
  \BibitemOpen
  \bibfield  {author} {\bibinfo {author} {\bibfnamefont {M.}~\bibnamefont
  {He}}, \bibinfo {author} {\bibfnamefont {J.}~\bibnamefont {Cai}}, \bibinfo
  {author} {\bibfnamefont {H.}~\bibnamefont {Zheng}}, \bibinfo {author}
  {\bibfnamefont {E.}~\bibnamefont {Seewald}}, \bibinfo {author} {\bibfnamefont
  {T.}~\bibnamefont {Taniguchi}}, \bibinfo {author} {\bibfnamefont
  {K.}~\bibnamefont {Watanabe}}, \bibinfo {author} {\bibfnamefont
  {J.}~\bibnamefont {Yan}}, \bibinfo {author} {\bibfnamefont {M.}~\bibnamefont
  {Yankowitz}}, \bibinfo {author} {\bibfnamefont {A.}~\bibnamefont
  {Pasupathy}}, \bibinfo {author} {\bibfnamefont {W.}~\bibnamefont {Yao}},\
  and\ \bibinfo {author} {\bibfnamefont {X.}~\bibnamefont {Xu}},\ }\bibfield
  {title} {\bibinfo {title} {Dynamically tunable moiré exciton rydberg states
  in a monolayer semiconductor on twisted bilayer graphene},\ }\href
  {https://doi.org/10.1038/s41563-023-01713-y} {\bibfield  {journal} {\bibinfo
  {journal} {Nature Materials}\ }\textbf {\bibinfo {volume} {23}},\ \bibinfo
  {pages} {224–229} (\bibinfo {year} {2024})}\BibitemShut {NoStop}%
\bibitem [{\citenamefont {Castellanos‐Gomez}\ \emph
  {et~al.}(2011)\citenamefont {Castellanos‐Gomez}, \citenamefont {Wojtaszek},
  \citenamefont {Tombros}, \citenamefont {Agraït}, \citenamefont {van Wees},\
  and\ \citenamefont {Rubio‐Bollinger}}]{Castellanos2011}%
  \BibitemOpen
  \bibfield  {author} {\bibinfo {author} {\bibfnamefont {A.}~\bibnamefont
  {Castellanos‐Gomez}}, \bibinfo {author} {\bibfnamefont {M.}~\bibnamefont
  {Wojtaszek}}, \bibinfo {author} {\bibfnamefont {N.}~\bibnamefont {Tombros}},
  \bibinfo {author} {\bibfnamefont {N.}~\bibnamefont {Agraït}}, \bibinfo
  {author} {\bibfnamefont {B.~J.}\ \bibnamefont {van Wees}},\ and\ \bibinfo
  {author} {\bibfnamefont {G.}~\bibnamefont {Rubio‐Bollinger}},\ }\bibfield
  {title} {\bibinfo {title} {Atomically thin mica flakes and their application
  as ultrathin insulating substrates for graphene},\ }\href
  {https://doi.org/10.1002/smll.201100733} {\bibfield  {journal} {\bibinfo
  {journal} {Small}\ }\textbf {\bibinfo {volume} {7}},\ \bibinfo {pages}
  {2491–2497} (\bibinfo {year} {2011})}\BibitemShut {NoStop}%
\bibitem [{\citenamefont {Hayrapetyan}\ \emph {et~al.}(2026)\citenamefont
  {Hayrapetyan}, \citenamefont {Sargsyan}, \citenamefont {Karakhanyan},
  \citenamefont {Khachatryan}, \citenamefont {Levonyan}, \citenamefont
  {Litvinov}, \citenamefont {Koperski}, \citenamefont {Margaryan},
  \citenamefont {Šiškins}, \citenamefont {Novoselov},\ and\ \citenamefont
  {Ghazaryan}}]{hay2026}%
  \BibitemOpen
  \bibfield  {author} {\bibinfo {author} {\bibfnamefont {M.}~\bibnamefont
  {Hayrapetyan}}, \bibinfo {author} {\bibfnamefont {M.}~\bibnamefont
  {Sargsyan}}, \bibinfo {author} {\bibfnamefont {D.}~\bibnamefont
  {Karakhanyan}}, \bibinfo {author} {\bibfnamefont {A.}~\bibnamefont
  {Khachatryan}}, \bibinfo {author} {\bibfnamefont {M.}~\bibnamefont
  {Levonyan}}, \bibinfo {author} {\bibfnamefont {D.}~\bibnamefont {Litvinov}},
  \bibinfo {author} {\bibfnamefont {M.}~\bibnamefont {Koperski}}, \bibinfo
  {author} {\bibfnamefont {A.}~\bibnamefont {Margaryan}}, \bibinfo {author}
  {\bibfnamefont {M.}~\bibnamefont {Šiškins}}, \bibinfo {author}
  {\bibfnamefont {K.~S.}\ \bibnamefont {Novoselov}},\ and\ \bibinfo {author}
  {\bibfnamefont {D.~A.}\ \bibnamefont {Ghazaryan}},\ }\href
  {https://arxiv.org/abs/2604.20023} {\bibinfo {title} {Broadband dielectric
  permittivity tensor of muscovite for next-generation all van der waals
  photonic components}} (\bibinfo {year} {2026}),\ \Eprint
  {https://arxiv.org/abs/2604.20023} {arXiv:2604.20023 [physics.optics]}
  \BibitemShut {NoStop}%
\end{thebibliography}

%

\end{document}